\documentclass{article}
\usepackage{spconf,amsmath,graphicx,url}

\title{Multilingual Alzheimer's Dementia Recognition through
  Spontaneous Speech: a Signal Processing Grand Challenge}
\name{\parbox{.65\linewidth}{\centering Saturnino
    Luz$^1$, Fasih Haider$^1$, Davida Fromm$^2$,  Ioulietta Lazarou$^3$, Ioannis Kompatsiaris$^3$ and Brian MacWhinney$^2$}}
\address{$^1$ Usher Institute, Edinburgh Medical School, The University of
  Edinburgh, UK\\
  $^2$ Department of Psychology, Carnegie Mellon University, USA \\
  $^3$ Information Technologies Institute, CERTH, Thermi-Thessaloniki, Greece
}
\begin{document}
%
\maketitle
\begin{abstract}
  This Signal Processing Grand Challenge (SPGC) targets a difficult
  automatic prediction problem of societal and medical relevance,
  namely, the detection of Alzheimer's Dementia (AD). Participants
  were invited to employ signal processing and machine learning
  methods to create predictive models based on spontaneous speech
  data.  The Challenge has been designed to assess the extent to which
  predictive models built based on speech in one language (English)
  generalise to another language (Greek). To the best of our knowledge
  no work has investigated acoustic features of the speech signal in
  multilingual AD detection. Our baseline system used conventional
  machine learning algorithms with Active Data Representation of
  acoustic features, achieving accuracy of 73.91\% on AD detection,
  and 4.95 root mean squared error on cognitive score prediction.
\end{abstract}
\vspace{-.5em}
\section{Introduction}
\label{sec:intro}
\vspace{-.5em}

Dementia is a category of neurodegenerative diseases that entail a
long-term and usually gradual decrease of cognitive functioning. As
cost-effective and accurate biomarkers of neurodegeneration have been
sought in the field of dementia research, speech-based ``digital
biomarkers'' have emerged as a promising possibility.  While there has
been much interest in automated methods for cognitive impairment
detection through speech by the signal processing and machine learning
communities \cite{bib:DelaFuenteRichieLuz2020JAD}, most of the
proposed approaches have not investigated which speech features can be
generalised and transferred across languages for AD prediction, and to
the best of our knowledge no work has investigated acoustic features
of speech in multilingual AD detection. This SPGC, ``ADReSS-M:
Multilingual Alzheimer's Dementia Recognition through Spontaneous
Speech'' targets this issue by defining prediction tasks whereby
participants train their models on English speech data and assess
their models' performance on spoken Greek data. The models submitted
to the challenge focus on acoustic or linguistic features of the
speech signal whose predictive power is preserved across languages.

This SPGC aims to provide a platform for contributions and discussions
on applying signal processing and machine learning methods for
multilingual AD recognition, and stimulate the discussion of machine
learning architectures, novel signal processing features, feature
selection and extraction methods, and other topics of interest to the
growing community of researchers interested in investigating the
connections between speech and dementia.

\vspace{-.5em}
\section{The prediction tasks}
\label{sec:tasks}
\vspace{-.3em}

The ADReSS-M challenge consists of the following tasks: (1) a
classification task, where the model will aim to distinguish healthy
control speech from AD/MCI speech, and (2) an MMSE score prediction
(regression) task, where you create a model to infer the speaker's
Mini Mental Status Examination (MMSE) score based on speech data.
Participants could choose to do one or both tasks. They were
provided with a training set and, two weeks prior to the
paper submission deadline, with test sets on which to test their
models. Up to five sets of results were allowed for scoring for each
task per participant.  All attempts had to be submitted together.

\vspace{-.5em}
\subsection{The data sets}
\label{sec:data-set}
\vspace{-.5em}

This SPGC data sets were made available through
DementiaBank\footnote{\url{https://dementia.talkbank.org/}}, upon
request.  The training dataset consists of spontaneous speech samples
corresponding to audio recordings of picture descriptions produced by
cognitively normal subjects and patients with an AD diagnosis, who
were asked to describe the Cookie Theft picture from the Boston
Diagnostic Aphasia Examination
test\cite{bib:BeckerEtAllPittCorpus94}. The participants were speakers
of English. The test set consists of spontaneous speech descriptions
of a different picture, in Greek. The recordings were made in one of
these languages. Participants were initially allowed access only to
the training data (in English) and some sample Greek data (8
recordings) for development purposes.

The Greek recordings assess participants' verbal fluency and mood
using a picture that the participant describes while looking at
it. The assessor first shows the participant a picture representing a
lion lying with a cub in the dessert while eating. The assessor then
asks the participants to give a verbal description of the picture in a
few sentences.

The training dataset was balanced with respect to age and gender in
order to eliminate potential confounding and bias. As we employed a
propensity score approach to matching we did not need to adjust for
education, as it correlates with age and gender, which suffice as an
admissible for adjustment (see \cite[pp 348-352]{bib:Pearl00causal}).
The dataset was checked for matching according to scores defined in
terms of the probability of an instance being treated as AD given
covariates age and gender estimated through logistic regression, and
matching instances were selected. All standardized mean differences
for the covariates were below 0.1 and all standardized mean
differences for squares and two-way interactions between covariates
were below 0.15, indicating adequate balance for those
covariates.

\vspace{-.5em}
\subsection{Evaluation}
\label{sec:evaluation}
\vspace{-.3em}

The classification task is evaluated in terms of accuracy,
specificity, sensitivity and $F_1$ scores. For the regression task
(MMSE prediction), the metrics used are the coefficient of
determination and root mean squared error. The ranking of submissions
is based on accuracy scores for the classification task (task 1), and
on RMSE scores for the MMSE score regression task (task 2).

\vspace{-.5em}
\section{Baseline models}
\label{sec:baseline-models}
\vspace{-.5em}

First we normalised the volume of audio files using ffmpeg' EBU R128
scanner filter.  A sliding window of 1 s, with no overlap, was then
applied to the audio, and eGeMAPS features were extracted over these
frames. The eGeMAPS feature set \cite{bib:EybenSchererEtAl16gg} is a
basic set of acoustic features designed to detect physiological
changes in voice production. It contains the F0 semitone, loudness,
spectral flux, MFCC, jitter, shimmer, F1, F2, F3, alpha ratio,
Hammarberg index and slope V0 features, as well as their most common
statistical functionals, totalling 88 features per frame. Given the
eGeMAPS features, we applied the active data representation method
(ADR) \cite{bib:HaiderFuenteLuz20aspacf} to generate a frame level
acoustic representation for each audio recording. The ADR method has
been used previously to generate large scale time-series data
representation. It employs self-organising mapping to cluster the
original acoustic features and then computes second-order features
over these clusters to extract new features
\cite{bib:HaiderFuenteLuz20aspacf}. This method is entirely automatic
in that no speech segmentation or diarisation information is provided
to the algorithm.

For task 1, we employed a Na\"{i}ve Bayes classifier with kernel
smoothing estimation. The ADR for feature extraction was optimised
using a grid search ($C= {5,10,\textbf{15},20,25} $).  We achieved
accuracies of 75.00\% and \textbf{73.91\%} on sample and test data
respectively using 15+2 ADR, age and gender features per recording. On
the test set, specificity was 79.2\%, precision was 75\%,
sensitivity was 68.2\%, and $F_1$ was 71.4\%. The feature to training
audio ratio was 19:237.

For the MMSE regression task (task 2), we employed a support vector
machine (SVM) model with a RBF kernel with box constraint of 1,
and sequential minimal optimization solver. The ADR for feature
extraction was optimised using a grid search
($C= {5,10,15,20,\textbf{25}}$).  This model achieved a root mean
squared error (RMSE) of 3.887 ($r= 0.273$) and \textbf{4.955}
($r= 0.348$) on sample and test data respectively using 25+2 ADR, age
and gender features per recording. The feature to training audio
recordings ratio was also 29:237.

\section{Conclusion}

Spontaneous speech analysis has the potential to enable novel
applications for speech technology in longitudinal, unobtrusive
monitoring of cognitive health.  By focusing on AD recognition using
spontaneous speech, this SPGC investigates an alternative to
neuropsychological and clinical evaluation approaches to AD detection
and cognitive assessment. Furthermore, the multilingual setting
provided by this SPGC allows the investigation of features that might
generalise across languages, extending the applicability of the
models.  In keeping with the objectives of AD prediction evaluation,
the ADReSS-M challenge provides a statistically matched data set so
as to mitigate common biases often overlooked in evaluations of AD
detection methods, including repeated occurrences of speech from the
same participant, variations in audio quality, and imbalances of
gender, age and educational level. We hope this might serve as a
benchmark for future research on multilingual AD assessment.

\let\oldbibliography\thebibliography
\renewcommand{\thebibliography}[1]{%
  \oldbibliography{#1}%
  \setlength{\itemsep}{2pt}%
}
\bibliographystyle{IEEEbib-abbrev}
\bibliography{madress}

\end{document}